\documentclass[conference]{IEEEtran}

\IEEEoverridecommandlockouts
\ifCLASSINFOpdf
  \usepackage[pdftex]{graphicx}
  \graphicspath{{../pdf/}{../jpeg/}}
\else
  \usepackage[dvips]{graphicx}
  \graphicspath{{../eps/}}
  \DeclareGraphicsExtensions{.eps}
\fi

\usepackage{subfigure} 
\usepackage{balance}

\begin{document}
%
\title{Unveiling Architecture Documentation: \\ Brazilian Stakeholders in 
Perspective}

\author{
  \IEEEauthorblockN{
  	Crescencio Rodrigues Lima Neto$^{1,2}$, 
  	Christina von Flach Chavez$^{1}$, Eduardo Santana de Almeida$^{1}$}
  	
  	\IEEEauthorblockN{
  	Dominik Rost$^{3}$,
  	Matthias Naab$^{3}$}\\
 
  \IEEEauthorblockN{
  	$^1$Computer Science Department - Federal University of Bahia (DCC/UFBA)}
   \IEEEauthorblockN{
  	$^2$Federal Institute of Bahia (IFBA)}
  	\IEEEauthorblockN{
  	$^3$Fraunhofer IESE}
   {\tt\{crescencio, flach, esa\}@dcc.ufba.br,} \\ {\tt\{dominik.rost,
   matthias.naab\}@iese.fraunhofer.de}
   \thanks{TR-PGCOMP-002/2015. Technical Report. Computer Science Graduate
   Program. Federal University of Bahia.}
 
   }

\maketitle

\begin{abstract}

Over the years, software architecture has become a established discipline, both
in academia and industry, and the interest on software architecture
documentation has increased. In this context, the improvement of methods, tools,
and techniques around architecture documentation is of paramount importance. We
conducted a survey with 147 industrial participants (31 from Brazil), analyzing
their current problems and future wishes. We identified that Brazilian
stakeholders need updated architecture documents with the right information.
Finally, the automation of some parts of the documentation will reduce the
effort during the creation of the documents. But first, is necessary to change
the culture of the stakeholders. They have to participate actively in the
architecture documents creation.

\end{abstract}







%



\IEEEpeerreviewmaketitle

\section{Introduction}

\label{sec:introduction}

Making software architecture explicit and persistent is a key factor for using
the potential it offers as an enabler for efficient and effective software
development, specially in scenarios of increasing system size and complexity,
and globally distributed development teams. This is reflected by the fact that
almost all comprehensive approaches for software architecture also address
documentation \cite{Garlan2010,Rozanski2005}, by the existence of a standard for
the description of software architectures \cite{isoiec2011}, and the concern
with architectural knowledge management \cite{Farenhorst}.

Nevertheless, many industrial organizations still have issues with respect to
software architecture documentation, that range from the absence of any
architectural documentation in place to the inability of leveraging the
potential that lies in existing documentation. As consequence, software
development may suffer from growing communication and alignment efforts, and
potential architecture erosion problems during evolution.

As a foundation for the improvement of methods and tools around software
architecture documentation, we conducted a survey with industrial participants,
investigating their current problems and wishes for the future, mainly focusing
on the developers point of view \cite{Rost2013}. 

We contacted 92 IT organizations from Europe, Asia, North and South America. A
total of 147 participants, from different countries (such as Germany, Finland,
Japan, USA, Sweeden and Brazil), answered the survey and were included in the
data analysis.

Our main findings were that architecture documentation should become up-to-date
and consistent in order to better serve the stakeholders's needs, and that they
demanded for more specific architecture documentation, targeted at their
concrete context and tasks.

In this work, we present a more detailed analysis on the answers provided by
Brazilian stakeholders. We depart from our previous work \cite{Rost2013} to
provide a preliminary characterization on the state-of-practice on architecture
documentation in Brazil. And therefore, create a basis for future improvement of
methods and tools for architecture documentation in Brazil, to make development
more efficient and effective.

We defined the main goal of this study according to the GQM template
\cite{Wohlin2000} as:

\emph{``Characterizing the current situation and improvement potential of
software architecture documentation with respect to architectural information
and its representation from the perspective of Brazilian developers in industry
as the basis for developing practically applicable methods and tools to make
implementation work more efficient and effective.''}

Our focus is Brazilian software developers, more specifically, a subset from
those software developers surveyed in \cite{Rost2013}. While methods and tools
might target architects in the documents' creation, in this study, we asked
stakeholders about their view as users of the documentation.

In this study, we refined the four research questions used in \cite{Rost2013}:

\begin{itemize}

\item \emph{RQ1: Which architectural information do stakeholders currently
receive for support the activities and which problems do they perceive?}

\item \emph{RQ2: Which representation of architectural information do
stakeholders currently receive for support the activities and which problems do
they perceive?}

\item \emph{RQ3: Which architectural information would stakeholders like to get
for their activities?}

\item \emph{RQ4: Which representation of architectural information would
stakeholders like to get for their activities?}

\end{itemize}

The remainder of this paper is structured as follows. Section
\ref{sec:methodology} presents the research methodology. Section
\ref{sec:results} presents the results. Section \ref{sec:discussion} presents
the main findings, discussion, and threats to the validity. Section
\ref{sec:conclusion} concludes the paper and presents future research.

 
 
 
 


\section{Research Methodology}\label{sec:methodology}
We followed the activities described in Kitchenham et al. approach
\cite{Kitchenham2008}: setting the objectives, survey design, developing and
evaluating the survey instrument (online questionnaire), conducting the survey,
and analyzing the data.

\subsection{Planning the Survey}
The focus group for the survey was software stakeholders in industry. According
to \cite{Garlan2010}, a stakeholder of an architecture is someone who has a
vested interest in it.

Thereby, it was not important whether they actually had software architecture
documentation available in their work, because asking them about their wishes
for the future was possible in either way. 

To distribute information about the survey we decided to use email, however we
did not want to just contact random software companies. To increase the chances
for a higher response rate, we compiled a list of fitting past and current
contacts from industry. 

As we typically have only one or two contact persons, we contacted them directly
and asked to distribute the information about the survey internally to software
stakeholders in their organization. In this way, we contacted IT organizations
(25 from Brazil), with four to around 100,000 employees.

\subsection{Designing and Conducting the Survey}
\begin{figure}[htp]
\centering
\includegraphics[width=3.6in]{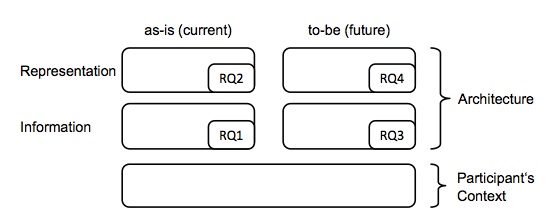}
\caption{Structure of survey questions and relationship \cite{Rost2013}}
\label{fig:rq}
\end{figure}

The research questions provided the framework for derivation of our survey
questions. Figure \ref{fig:rq} presents the resulting survey questions structure
as a matrix. The key distinction between the as-is situation for the participant
and wishes for a to-be situation related to architecture documentation.
Moreover, we asked for information about the participants' background. 

\begin{figure*}[htp]
\centering
\includegraphics[width=7in]{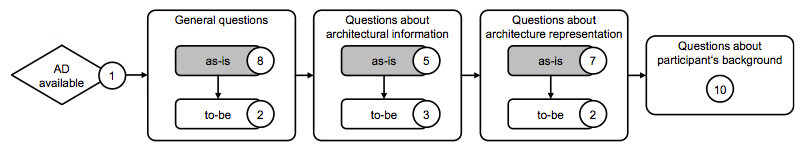}

\caption{Flow of questions in the survey adapted from \cite{Rost2013}. Grey
blocks are asked in case of architecture documentation available. The
circles indicate the number of questions.}
\label{fig:flow}
\end{figure*}

Figure \ref{fig:flow} presents the survey questions flow. It starts with a
question about the availability of architecture documentation for the
participants. This question has an impact on the further survey questions flow:
only if a participant indicates that he has architecture documentation
available, the questions about the as-is situation are asked, otherwise they are
not visible for the participant.

The survey main part are three pages of questions, each visually separated into
a set for the as-is situation and a set for the to-be situation: first, we asked
general questions about architecture documentation, not differentiating between
the information aspects and their representation.
Second, we asked questions with a focus on architectural information in
architecture documentation. Third, we asked questions about the architecture
information representation. Finally, we asked questions about the participants'
background.

We had two types of questions: first, question with a fixed set of answers,
partially single and partially multi selection ones. Second, there were
questions with a free text answer. Moreover, we created an online
questionnaire containing 42 questions. We conducted the survey in the period
from December 1st, 2012 to January 31st, 2013.

\subsection{Analyzing the Data}
Only a subset of the participants who started the survey actually finished it.
We considered the survey as finished when the participants finished and submitted
it. For the analysis and evaluation, when we talk about participants we refer to
the ones having finished the survey.

In total, 147 stakeholders (31 Brazilians) from different countries (such as
Germany, Brazil, Finland, Japan, USA, and Sweeden) participated  and have been
included in the data analysis.

To analyze the Brazilian perspective, we separate the Brazilian stakeholders
(BRA) from the other (N-BRA) participants. Nevertheless, we did not have a
complete data set for each question, as not all questions were mandatory. That
is, for each question the sample varies.

As described, we asked for this availability and excluded the questions about
current architecture documentation if there is none. Not all participants had
architecture documentation available for their tasks. Thus, for the questions
about current architecture documentation we have only answers of a subset of the
participants.

For questions with fixed answers we counted the results in the analysis. For the
evaluation of free text results, we grouped the answers into coherent categories
with a name chosen by us as perceived meaningful to cover the full range of
answers. 

Then we also aligned these answer categories across questions where it
was meaningful. Finally, we analyzed the survey data based on descriptive
statistics and hypothesis testing.


\section{Results}\label{sec:results}

In this section we analyzed all the survey participants answers.

\subsection{Overview of Survey Participants and their Context}
All participants work in industry and are related to software development. In
case of the Brazialian participants, they are affiliated to Brazilian or
multinational companies located in Brazil.

\begin{figure}[htp]
\centering
\includegraphics[width=3.4in]{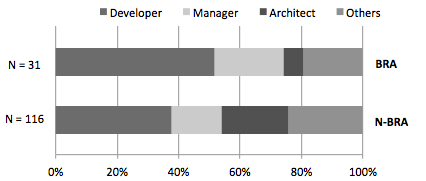}
\caption{Current position of the participants in their companies}
\label{fig:position}
\end{figure}

The survey aims at the development perspective on software architecture. Figure
\ref{fig:position} depicts the distribution of participants' occupational positions. 

We use the stakeholders classification proposed by \cite{Garlan2010}. Moreover,
we considered developer as a synonym for implementer. As suggested by Garlan,
the implementer is responsible for the development of specific elements according to
designs, requirements, and architecture. In other words, they have the same
meaning.

The survey initial purpose focused only on how developers used the
architecture documentation. Since the documentation is constantly used by
several stakeholders, we included software development stakeholders
such as architects, designers, testers, etc.

The Brazilian participant largest group is developers (52\%), followed by
managers (22\%) and architects (7\%). On the other hand, the participation of
non-Brazilian developers (38\%) and architects (22\%) was significant. The
participants' position was asked as free text, thus we grouped the answers into
the depicted categories.

In order to evaluate the participants professional experience in their position,
we asked them for the number of years working in this or a similar position. The
answers, grouped from an open question, were the following: 
\begin{itemize}
  \item 0 to 3 years: BRA -- 16\% and N-BRA -- 33\%;
  \item 4 to 7 years: BRA -- 52\% and N-BRA -- 33\%;
  \item 8 to 11 years: BRA -- 23\% and N-BRA -- 15\%;
  \item 12 to 15 years: BRA -- 3\% and N-BRA -- 12\%; and
  \item more than 15 years: BRA -- 6\% and N-BRA -- 13\%.  
\end{itemize}

\begin{figure}[htp]
\centering
\includegraphics[width=3.5in]{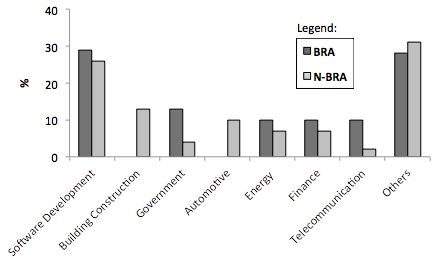}
\caption{Sectors of the participants' companies}
\label{fig:caracterization}
\end{figure}

In order to characterize the participants affiliation, we asked for the industry
sectors they work in (see Figure \ref{fig:caracterization}). Most participants
(both Brazilian and non-Brazilian) work for companies that have software
development for multiple industries (27\%) as their main business. Other
Brazilian strong sectors in our survey are government (13\%), energy (10\%),
finance (10\%), and telecommunication (10\%). The non-Brazilian strong sectors
are building construction management (13\%) and automotive (10\%)

While the survey in general was anonymous, we asked the participants at the end
whether they agree with publishing their company’s name in the study.
Participating companies included among others: Recife Center for Advanced
Studies and Systems (C.E.S.A.R), Accenture, Ogilvy \& Mather, Serpro, Dataprev,
ThoughtWorks.

We identified differences in the number of people in the companies contributing
to software development. The answers were the following:


\begin{itemize}
  \item none: BRA -- 0\% and N-BRA -- 15\%;
  \item less than 100: BRA -- 39\% and N-BRA -- 44\%;
  \item 100 to 1000: BRA -- 26\% and N-BRA -- 34\%;
  \item 1001 to 5000: BRA -- 22\% and N-BRA -- 3\%; and,
  \item more than 5001: BRA -- 13\% and N-BRA -- 4\%;
\end{itemize}

The majority (BRA -- 58\% and N-BRA -- 45\%) of the participants companies
develops software according to a combination of agile and conventional
development processes. In the case of Brazilian stakeholders, 23\% work with
agile development processes, only 6\% work with conventional development
processes, and 13\% do not use a structured development process at all. The
numbers are similiar for non-Brazilian participants, 32\% work with agile
development processes, and only 7\% work with conventional development
processes.

Furthermore, we asked them to rate the product size they are contributing to. In
order to simplify the answering, the participants defined the product size
estimate, were the following:

\begin{itemize}
  \item very large: BRA -- 19\% and N-BRA -- 22\%;
  \item large: BRA -- 26\% and N-BRA -- 32\%;
  \item medium-size: BRA -- 29\% and N-BRA -- 34\%
  \item small: BRA -- 23\% and N-BRA -- 9\%; and,
  \item very small: BRA -- 3\% and N-BRA -- 4\%.
\end{itemize}

Finally, we identified that 87\% of the Brazilian stakeholders that work
with large product size, also work with a combination of agile and conventional
development process.

\subsection{Architectural Information: The as-is Situation}

In the general questions, we asked \emph{``What do you consider as the main
problems with the architecture documentation you work with?''} and received with
respect to architectural information the following most frequent questions:
\begin{itemize}
  \item Outdated architecture documentation;
  \item Inadequate level of granularity;
  \item Implementation not in sync with architecture; and 
  \item Unnecessary information.
\end{itemize}

\begin{figure*}[htp]
\centering
\includegraphics[width=7in]{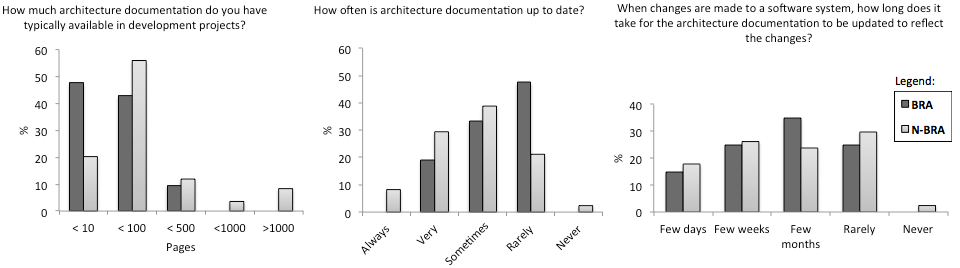}
\caption{Amount and up-to-dateness of architecture documentation}
\label{fig:up-to-date}
\end{figure*}

We asked about the amount of architecture documentation available and its
up-to-dateness. The results are depicted in Figure \ref{fig:up-to-date}. This
also confirms that architecture documentation is often not up-to-date and if at
all updated with a strong delay. This supports the findings reported in
\cite{Lethbridge2003}.

The stakeholders' majority has access to architecture documentation with
less than 100 pages (BRA -- 43\% and N-BRA -- 56\%). It is worthwhile to mention
that almost half of the Brazilian participants (48\%) has, only, 10 pages of
architecture documentation available in their development projects.

Moreover, 56\% of the participants (both Brazilian and non-Brazilian) deal with
documents up to 10 and 100 pages. Addressing this concern, 83\% of them document
the architecture of very large (22\%), large (29\%) and medium (33\%) products.
In other words, stakeholders that work with larger product size, have
insufficient documentation available (fewer pages). 



About the architecture documentation up-to-dateness, on the one hand, only 8\%
of the participants (non-Brazilian stakeholders) access documentation that is
always up-to-date, on the other, 48\% of the Brazilian stakeholders access
documentation that is rarely up-to-date.


We also identified that, in the case of the Brazilian participants, 70\% of the
documentation up to 10 pages are rarely up-to-date. Therefore, two facts were
observed: (i) The project architecture did not change and (ii) The architectural
decision making is not documented at all.


Finally, we asked the participants \emph{``When changes are made to a
software system, how long does it take for architecture documentation to be
updated to reflect the changes?''} The answers were the following:

\begin{itemize}
  \item Few days: BRA -- 15\% and N-BRA -- 18\%;
  \item Few weeks: BRA -- 25\% and N-BRA -- 26\%;
  \item Few months: BRA -- 35\% and N-BRA -- 24\%;
  \item Rarely: BRA -- 25\% and N-BRA -- 30\%; and
  \item Never: BRA -- 0\% and N-BRA -- 2\%.
\end{itemize}

When changes occur in the project documents up to 100 pages, they are rarely
updated (53\%). In documentation up to 10 pages, 27\% of the documents are
rarely up-to-date.

\begin{figure}[htp]
\centering
\includegraphics[width=3.5in]{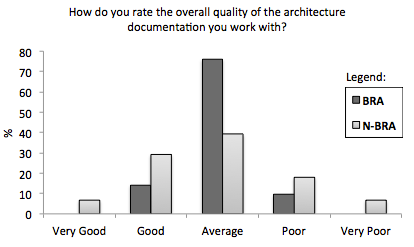}
\caption{Architecture documentation quality}
\label{fig:quality}
\end{figure}

Regarding the architecture documentation quality, the Figure \ref{fig:quality}
presents how the stakeholders rate the architecture documentation quality.

Although access outdated documents with a small number of pages, the
participants majority considers the quality of the document as average (BRA --
76\% and N-BRA -- 39\%). Furthermore, 96\% of the documentation that rarely
reflects the changes are qualified as poor (37\%) and average (60\%).



\begin{figure}[t]
\centering
\includegraphics[width=3.4in]{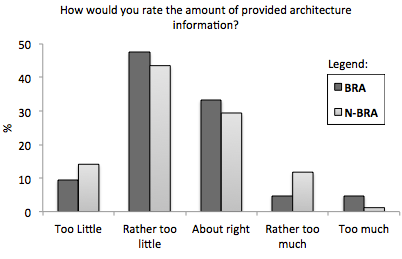}
\caption{Adequate amount of architecture documentation}
\label{fig:adequacy_1}
\end{figure}

We asked the participants to rate their perceived adequacy of the amount of
architecture information provided (as it can be seen in Figure
\ref{fig:adequacy_1}). A tendency can be observed that there is rather too
little architectural information available (BRA -- 48\% and N-BRA -- 44\%). Some
participants agree that there is also necessary information but most
participants see unnecessary information provided.

\begin{figure}[htp]
\centering
\includegraphics[width=3.5in]{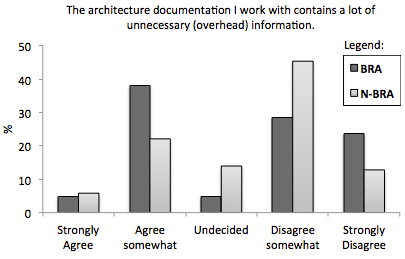}
\caption{Amount of unnecessary (overhead) information}
\label{fig:adequacy_2}
\end{figure}

Figure \ref{fig:adequacy_2} presents the amount of unnecessary (overhead)
information in the architecture documentation. While most Brazilian participants
(38\%) work with documentation that contains a lot of unnecessary information,
this is not an issue for most non-Brazilian participants (45\%).

Moreover, it is worthwhile to mention that part of these non-Brazilian
stakeholders (49\%), work with documentation up to 100 pages that are Very Often
(38\%) or Sometimes (38\%) updated.

\begin{figure*}[htp]
\centering
\includegraphics[width=7in]{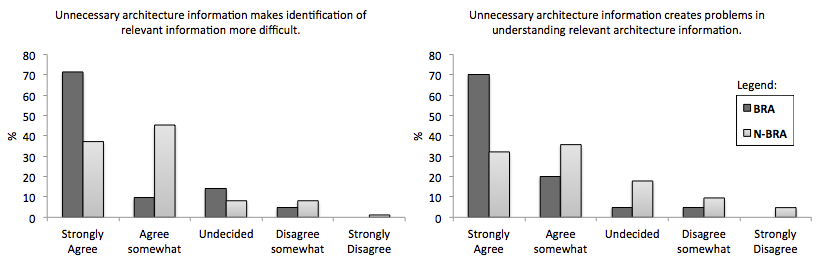}
\caption{Unnecessary information in the architecture documentation}
\label{fig:unecessary}
\end{figure*}

Figure \ref{fig:unecessary} shows that most of participants confirms that
unnecessary information in the architecture documentation interferes on the
understanding of architecture information making identification of relevant
information more difficult.

We also asked the participants \emph{``Do you see any other problems or
consequences because of unnecessary architecture information?''} the
following most frequent questions:

\begin{itemize}
  \item Unnecessary information make the stakeholder pay attention in points
  that are not relevant;
  \item It make the documents long and difficult to read;  
  \item People tend to loose interest if given too complex or blurred
  information to handle; and
  \item Using time to create and update unnecessary architecture information is
  obviously time away from something more important.
\end{itemize}

\subsection{Representation of the Architectural Information: The as-is
Situation}

In the general questions, we asked \emph{``What do you consider as the main
problems with the architecture documentation you work with?''} and received with
respect to representation of architectural information, the following most
frequent questions:

\begin{itemize}
  \item Inconsistencies and missing structure;
  \item Information scattered across documents; and
  \item Missing traceability to other artifacts.
\end{itemize}

In order to get some insights into the problems with the representation of
architecture information, we asked the question described in Table
\ref{table:problems}.

\begin{table}[h]
\centering	
\caption{What problems do you see in the way how architecture information is described?}
	\begin{tabular}{lcc}
	    \hline
	    	Answer Category & BRA (\%) & N-BRA (\%) \\
	    \hline
	        Unnecessary information & 40 & 3 \\
	        Documents not up-to-date & 30 & 0 \\
	        Missing details in the written description & 20 & 3 \\
	        Targets too many groups & 10 & 13 \\
	        Missing common formats and structures & 0 & 20 \\
	        Other & 0 & 60 \\
	    \hline
	\end{tabular} 
\label{table:problems}
\end{table}

We asked about the main formats in which architecture documentation is provided.
Architecture documentation is mostly provided as electronic documents (BRA
-- 68\% and N-BRA -- 65\%), model files (BRA -- 35\% and N-BRA -- 38\%), and web
pages (BRA -- 23\% and N-BRA -- 36\%).



In contrast to what Petre states in \cite{Petre2013}, UML is the key notation
for the description of software architecture in our survey (BRA -- 55\% and N-BRA
-- 58\%). In Brazil, ADLs are not used at all and only 6\% of the non-Brazlian
participants worked with it.

\begin{figure}[htp]
\centering
\includegraphics[width=3.5in]{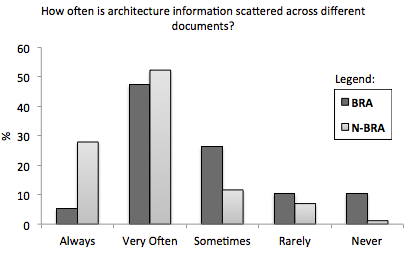}
\caption{Architecture information scattered across documents}
\label{fig:scattering}
\end{figure}

Moreover, the architecture information is often described in Natural Language
(BRA -- 55\% and N-BRA -- 71\%). The participants also used other forms of
representation such as visio, pseudo-code, and informal diagrams. Another result
is that architecture information is scattered across documents (see Figure
\ref{fig:scattering}).

\begin{figure}[htp]
\centering
\includegraphics[width=3.5in]{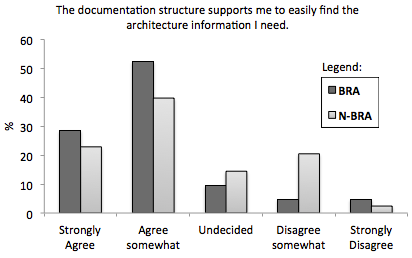}
\caption{Perceived representation adequacy of architecture information}
\label{fig:representation_1}
\end{figure}

We asked the participants how they perceive the support of their architecture
documentation to find specific information and to conduct their development
tasks. Figure \ref{fig:representation_1} presents a tendency that the
participants perceive the representation as adequate. To refine the insights about the
problems finding the needed architecture information, we asked the question
described in Table \ref{table:finding}.

\begin{table}[h]
\centering	
\caption{What problems do you see in terms of finding the architecture information you need?}
	\begin{tabular}{lcc}
	    \hline
	    	Answer Category & BRA (\%) & N-BRA (\%) \\
	    \hline
		    Documents not up-to-date & 29 & 7 \\
		    Missing strong search functionality & 21 & 13 \\
		    Too much information & 14 & 4 \\
		    Missing information & 14 & 4 \\
		    Information scattered across documents & 7 & 18 \\
		    Missing traceability & 7 & 13 \\
		    Other & 7 & 18 \\		    
		    Missing clarity in structure & 0 & 24 \\
	    \hline
	\end{tabular} 
\label{table:finding}
\end{table}

\begin{figure}[htp]
\centering
\includegraphics[width=3.4in]{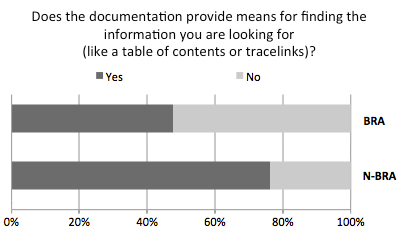}
\caption{Means providence for finding information}
\label{fig:Location}
\end{figure}

Figure \ref{fig:Location} presents that only 50\% of the Brazilian participants
work with documentation that provide means for finding the information they are
looking for.

We also asked the participants what means are included, the following most
frequent questions:
\begin{itemize}
  \item Table of Contents;
  \item Hyperlinks and Tracelinks;
  \item Search engine;
  \item Full-text search of wiki pages; and
  \item Tagging.
\end{itemize}

\begin{figure}[t]
\centering
\includegraphics[width=3.5in]{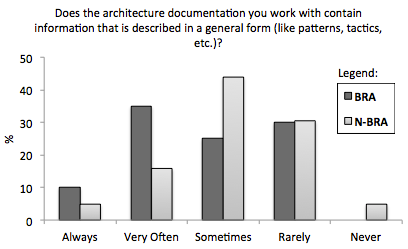}
\caption{Information described in general form}
\label{fig:adaptation}
\end{figure}

Figure \ref{fig:adaptation} presents how frequent the architecture documentation
is described in a general form. 35\% of the Brazilian participants work, Very
Often, with information described in patterns, tactics, etc. In contrast, 44\%
of the non-Brazilian participants work with this kind of information.

\begin{figure}[htp]
\centering
\includegraphics[width=3.5in]{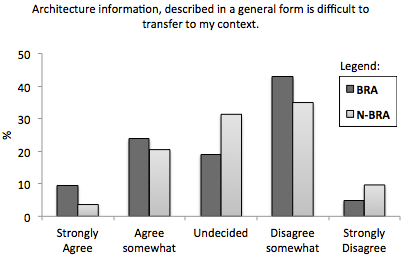}
\caption{Context transfer difficulty}
\label{fig:adaptationProblems}
\end{figure}

We asked the participants \emph{``Architecture Information, described in a
general form is difficult to tranfer to my context''} as it can be seen in
Figure \ref{fig:adaptationProblems}. Moreover, the participants' majority (BRA
-- 43\% and NBRA -- 35\%) disagree with the question.
 
\begin{figure}[htp]
\centering
\includegraphics[width=3.5in]{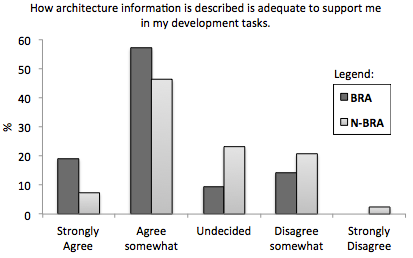}
\caption{Information adequacy to support development tasks}
\label{fig:representation_2}
\end{figure}

Finally, Figure \ref{fig:representation_2} presents the documentation adequacy
to support the participants' development tasks. Both, Brazilian participants
(57\%) as well as non-Brazilian participants (46\%) agree somewhat with the
question.

\subsection{Architectural Information: The to-be Situation}
In the general questions, we asked \emph{``What are your wishes in general for
the future of architecture documentation?''} and received with respect to
architectural information the following most frequent questions:

\begin{itemize}
  \item Up-to-date;
  \item In sync with implementation;
  \item Providing a system overview and the big picture; and
  \item Specific for stakeholders, tasks and contexts.
\end{itemize}


Furthermore, we asked the participants \emph{``What architecture information do
you need for best support of your development tasks?''} We identified that it is
more important to stakeholders to get an overall understanding of the complete
system, as well as detailed information on components in their scope together
with interfaces and relationships to other components. Table
\ref{table:bestsupport} presents an overview.


\begin{table}[h]
\centering	
\caption{What arch. information do you need for best support of your
development task?}
	\begin{tabular}{lcc}
	    \hline
	    	Answer Category & BRA (\%) & N-BRA (\%) \\
	    \hline
		    Components, interfaces, relationships & 47 & 33 \\
		    Big picture & 28 & 13 \\
		    Functional modularization & 15 & 7 \\
		    Others & 10 & 29 \\
		    Data model and data flow & 0 & 10 \\
		    Patterns and best practicies & 0 & 8 \\
	    \hline
	\end{tabular} 
\label{table:bestsupport}
\end{table}

\subsection{Representation of Architectural Information: The to-be Situation}
We asked the question \emph{``What are your wishes in general for the future of
architecture documentation?''} and got with respect to representation of
architectural information the following most frequent questions:

\begin{itemize}
  \item Easy creation, handling, updating, and maintenance;
  \item Integrated information, artifacts and tools;
  \item Readable and understandable; and
  \item Consistent, structured, and described.
\end{itemize}

Moreover, we also asked the participants \emph{``In what format should
architecture documentation be provided in the future?''} Table
\ref{table:format} presents an overview. It can be noted that webpages and
electronic documents are the wishes.  

\begin{table}[h]
\centering	
\caption{In what format should architecture documentation be provided in the future?}
	\begin{tabular}{lcc}
	    \hline
	    	Answer Category & BRA (\%) & N-BRA (\%)\\
	    \hline
		    Webpages & 41 & 16 \\
		    Electronic documents & 21 & 20 \\
		    Diagrams & 12 & 21 \\
		    UML & 9 & 18 \\
		    Wikis & 9 & 7 \\
		    Natural language & 4 & 17 \\
		    Other & 4 & 1 \\
	    \hline
	\end{tabular} 
\label{table:format}
\end{table}

Then we asked \emph{``What means should architecture documentation provide to
help you in finding the information you need?''} It can be observed that
stakeholders wish for interactive ways of working with architecture
documentation, where it is possible to search information in different ways and navigate
through hierarchical structures and related elements. Also mapping the
architecture to the implementation (source code) have been rated as important.
Table \ref{table:informationneed} presents an overview of the result data.

\begin{table}[h]
\centering	
\caption{What means should architecture documentation provide to help you in
finding the information you need?}
	\begin{tabular}{lcc}
	    \hline
	    	Answer Category & BRA (\%) & N-BRA (\%) \\
	    \hline
		    Interactive search functionality & 21 & 20 \\
		    Links and navigation & 21 & 19 \\
		    Mapping to implementation & 16 & 5 \\
		    Other & 16 & 27 \\		    
		    Traces to artifacts & 11 & 13 \\
		    Directories & 10 & 8 \\
		    Clear structure & 5 & 8 \\
	    \hline
	\end{tabular} 
\label{table:informationneed}
\end{table}

Finally, we asked how architecture should be described to make it useful for the
developer's implementation task. Table \ref{table:informationdescribed} presents
the results. It becomes evident that clarity and structure are of highest
importance, in diagrams and language.

\begin{table}[h]
\centering	
\caption{How should arch. information be described to make it more useful for your dev. tasks?}
	\begin{tabular}{lcc}
	    \hline
	    	Answer Category & BRA (\%) & N-BRA (\%) \\
	    \hline
		    Self-explaining, simple diagrams  & 50 & 17 \\
		    Clear, concise, uniform, consistent & 36 & 10 \\
		    Clear terminology and language & 14 & 8 \\
		    Other & 0 & 65 \\
	    \hline
	\end{tabular} 
\label{table:informationdescribed}
\end{table}

\subsection{Statistic Descriptive Analysis}
To identify relevant insights, we analyzed all the information produced by the
participants. The analysis was performed based on Shapiro-Wilk normality test
and Mann-Whitney-Wilcoxon \cite{Wohlin2000} (for non-parametric statistical
hypothesis test).

The Likert scale \cite{Jamieson2004} (used in some questions of the survey)
measures the extent to which a person agrees or disagrees with the question. The
most common scale is 1 to 5. For this reason the following scale were used:
\begin{itemize}
  \item Strongly Disagree = 1;
  \item Disagree somewhat = 2;
  \item Undecided = 3;
  \item Agree somewhat = 4; and
  \item Strongly Agree = 5.  
\end{itemize}

We selected five survey questions to analyze them:

\begin{itemize}
  \item \emph{Q1: The architecture documentation I work with contains a lot of
  unnecessary (overhead) information?}
  \item \emph{Q2: Unnecessary architecture information makes identification of
  relevant information more difficult?}
  \item \emph{Q3: The documentation structure supports me to easily find the
  architecture information I need?}
  \item \emph{Q4: Architecture information, described in general form is
  difficult to transfer to my context?}
  \item \emph{Q5: The way how architecture information is described is adequate
  to support me in my development tasks?}
\end{itemize}

\subsubsection{Hypothesis Formulation}
Based on the previously selected questions, we set up five hypotheses to
identify if the Brazilian participants (described below as $BRA$) answered the
questions differently from the non-Brazilian participants (described below as
$N-BRA$).

The Null Hypothesis ($H_{0n}$) considered that there was no difference in the
answers. The Null hypotheses are:

\begin{center} 
$
H_{0n}: \mu_{answers_{N-BRA}} = \mu_{answers_{BRA}}
$

\end{center}

Conversely, the Alternative Hypothesis ($H_{1n}$) stated the opposite values.
The alternative hypothesis determined that Brazilian participants (described
below as $BRA$) answer the questions differently from the non-Brazilian
participants. We herein define the set of alternative hypotheses, as follows:

\begin{center} 
$
H_{1n}: \mu_{answers_{N-BRA}} \neq \mu_{answers_{BRA}}
$

\end{center}

\subsubsection{Hypotheses Testing}

First, in order to reduce the dataset errors, we eliminated the outliers.
Second, we verified if the sample came from a normally distributed population
through Shapiro-Wilk test \cite{Wohlin2000}. Third, since Likert scale questions
do not possess a normal probability distribution, in other words, the samples
were not ``normal'', we used a non-parametric test (Mann-Whitney-Wilcoxon
\cite{Wohlin2000}) to analyze the hypothesis.

The tests are primarily presented for a significance level of 5\%. Table
\ref{tab:hypoTesting} presents the hypothesis testing results from the
selected survey questions. The results are detailed next.

\begin{table}[ht]
\centering	
\caption{Hypotheses Testing Results}
	\renewcommand{\arraystretch}{1.3}
	\label{tab:hypoTesting}
	\begin{tabular}{c c c}
	\hline
	\textbf{Question} & \textbf{p-value} & \textbf{Statistical Diference?}\\
	\hline
		Q1 & 0.878 & No \\
		Q2 & 0.031 & Yes \\
		Q3 & 0.192 & No \\
		Q4 & 0.641 & No \\
		Q5 & 0.046 & Yes \\
	\hline
	\end{tabular}
\end{table}

As regards the unnecessary information in the documentation (overhead), there
were no differences in the answers of Brazilian and non-Brazilian
participants. The Null Hypothesis \emph{$H_{01}$} cannot be rejected, since
there is no significant difference. The \emph{p-value} = 0.878, is higher than
5\%, which did not reject the null hypothesis, answering the Q1.

Regarding Q2 (Unnecessary information makes identification of relevant
information more difficult), the amount of Brazilian participants answers is
higher than the non-Brazilian, \emph{p-value} = 0.031. The \emph{p-value} is
smaller than the significance level, rejecting the null hypothesis \emph{$H_{02}$}.

Moreover, the Null Hypothesis \emph{$H_{03}$} and \emph{$H_{04}$} cannot
be rejected, since \emph{p-value} = 0.192 and 0.641 are higher than the
significance level the hypotheses cannot be rejected and no conclusion can be
drawn.

Therefore, in regard to Q3 (support to easily find architecture information) and
Q4 (information described in general form is difficult to transfer to their
context), there are no significant differences between the answers from
Brazilian and non-Brazilian participants.

Finally, regarding the Q5 (adequacy to support the stakeholders development
task), the \emph{p-value} = 0.046, is lower than 0.05, rejecting the Null Hypothesis
\emph{$H_{05}$}. In other words, there were differences in the answers of
Brazilian and non-Brazilian participants.

\section{Discussion}\label{sec:discussion}
In this section, we present the main findings, discuss the survey results, and
describe threats to validity.

\subsection{Main Findings}

This section describes the results of the survey. Please note that the results
of the general questions are consistently integrated into this structure. Since
this study focused on the perspective of the Brazilian participants, we
identified 5 main findings regarding this issue, which are summarized below:

\begin{enumerate}
  \item The Brazilian stakeholders are used to work small documents (less than
  100 pages). The majority of the documentation is provided with unnecessary
  and outdated information;
  \item The decision making is lost during the project development because
  there is no concern with traceability;
  \item The stakeholders (both Brazilian and non-Brazilian stakeholders) wish
  online and interactive documentation with a powerful search tool that return
  a more accurate result;
  \item The Brazilian stakeholders do not use formal ADLs to document the
  architecture; and finally, 
  \item The architecture documentation of most large companies (more than 1500
  employees) are small (less than 10 pages) and rarely updated.
  
  
%
  
\end{enumerate}

\subsection{Survey Results}

In accordance with \cite{Softex2012}, in Brazil, the majority of software
development companies is composed by small and medium-sized enterprises (Small
companies have up to 100 employees and medium-sized have up to 500 employees).
In many cases, there is no investment in documentation. As stated by Clements et
al. \cite{Garlan2010}, ``documentation is often treated as an afterthought''.
From this context, we revisit our main research questions and the responses we
received from the participants.
 
In regard to architectural information, we identified the same participants
concern to up-to-dateness as described in \cite{Rost2013}. The documents used by
Brazilian stakeholders are simple (due to the small number of pages) but they
need the description of the right information.

Regarding the representation of architecture information, the development of a
``online documentation'' was the main concern of stakeholders. They considered
the implementation of a website to allow the easy creation, handling, updating, and
maintenance of the architecture documentation. Moreover, all the information
about the architecture (including traceability and decision making) can be
concentrated in one place.

The automation of some parts of the architecture documentation is needed, but
how to do it once the stakeholders are reluctant to adopt ADLs and DSLs (Domain
Specific Languages)? The use of specific languages enables the connection of the
architecture documentation with the source code. The changes in the architecture
can be mapped in the source code and vice-versa.

As the statistic analysis confirms, the amount of unnecessary architecture
information makes identification of relevant information more difficult for the
Brazilian stakeholders. Furthermore, the information description is inadequate
to support their development tasks.

Finally, in the survey context, we identified that the ideal architecture
documentation has up to 100 pages sometimes updated and reflecting the source
code changes monthly. Moreover, the number of pages and update frequency are
influenced by the system complexity and, consequently, the product size.





%
%

\subsection{Threats to Validity}
According to Wohlin et al. (2000), it is important to consider the question of
validity already in the planning phase in order to anticipate possible threats
involved in the context of a survey. Following, we describe threats
to validity and limitations of the survey.

\subsubsection{Generalization of participants} This is an effect of having a
participant population not representative of the population we would like to
generalize to, i.e. the wrong people participating in the survey.
  
For this reason, we decided to invite the participants using e-mail, however we
did not want to just contact random software companies. To increase the chances
for a high response rate, we compiled a list of fitting past and current
customers and project partners from industry.
  
As we typically have only one or two contact persons, we contacted them directly
and asked to distribute the information about the survey internally to
stakeholders in their organization with the request to participate.
  
\subsubsection{Maturation} This is the effect that subjects react differently as
time passes. Some subjects can be affected negatively (feel bored or tired)
during the survey. In order to mitigate this boredom, the subjects were free to
choose the moment when they were comfortable. Even so, only 147 of the 350
participants completed the survey.
  
\subsubsection{Evaluation apprehension} Some people are afraid of being
evaluated \cite{Wohlin2000}. The participants could be afraid that the survey
affected their work. In order to avoid this risk, we explained that the
survey in general was anonymous. None of their personal information were
exposed. During the survey completion, the participants executed their
activities without pressure and without time restriction.

\subsubsection{Language comprehension} As this research was done in a
German-Brazilian cooperation with many participants from Germany and Brazil, we
offered the participants to choose their preferred languages such as German,
Portuguese and in addition, English.

\section{Concluding Remarks and Future Work}\label{sec:conclusion}
We conducted a study on the as-is situation of software architecture
documentation from the perspective of Brazilian stakeholders. We analyzed the
contributions from 147 (Brazilian and non-Brazilian) participants from industry
(31 of them from Brazil). 


The software architecture is a topic of interest of many companies. Learn how to
document the software architecture is essential in to produce high-quality
products, reduce costs and rework, and maintain the traceability
\cite{Garlan2010}.

We identified some improvement opportunities in \cite{Rost2013}, which were
extended with the Brazilian perspective in this work. The Brazilian stakeholders
concerns in not so different from international stakeholders concerns. For this
reason, some of our main findings are similar but with some particular
differences.

The stakeholders wish the creation of self-explaining and simple diagrams. They
need clear, concise, uniform, and consistent information. The architecture
documentation has to become up-to-date and interactive. The development of a
website combined with a powerful search tool is needed.

We believe that architecture documentation should be created for the user
understanding. Moreover, the automation of some parts of the documentation will
reduce the effort during the creation of the documents. But first, is necessary
to change the stakeholders ``culture''. They have to become active actors on the
creation of the architecture documentation.

Finally, it is necessary to document and update the ``right information'', in
the ``right moment''. Complex systems tend to require detailed architecture
documentation. On the one hand, companies with high employees turnover that
invest in documentation, reduce the costs when hiring a new employee.
On the other hand, when the architecture is constantly changing, it is relevant
to document the architectural decision making.





\subsection{Future Work}
We intend to extend this work to Product Line Architecture documentation. We
propose to investigate the differences from single system architecture
documentation to Product Line Architecture documentation.

In most cases of the Brazilian context, the necessity of architecture
documentation appears when the companies finish the code. For this reason, we
also propose the investigation of software architecture recovery techniques.


\section*{Acknowledgment}
This survey has been conducted in the context of the Fraunhofer Germany-Brazil
cooperation between Fraunhofer IESE and UFBA. We would like to thank all the
participants that contributed by answering our questions. This work was funded
by IFBA grants EDITAL Number BP003-04/2014/PRPGI.



\balance
\bibliographystyle{IEEEtran}
\bibliography{example}
%
%
%

\end{document}